# The application of precision time protocol on EAST timing system

Z. Zhang, B. Xiao, Z. Ji, Y. Wang, P. Wang

*Abstract*—The timing system focuses on synchronizing and coordinating each subsystem according to the trigger signals. The former timing system was based on commercial off-the-shelf devices and a set of synchronized optical network which was made up of several pairs of multi-mode fibers. The expensive PXI devices and inconvenient extension methods compel maintainers to upgrade the timing system to meet the ever increasing demands of the experiments. A new prototype timing slave node based on precision time protocol has been developed by using ARM STM32 platform. The proposed slave timing module is tested and results show that the synchronization accuracy between slave nodes is in sub-microsecond range.

## I. INTRODUCTION

THE Experimental Advanced Superconducting Tokamak (EAST) made an exciting advance in achieving a stable 101.2 second steady-state high confinement plasma, creating a new world record in long-pulse H-mode operation on July 3rd 2017. The timing system plays an important role during the discharge experiment to ensure the stable operation of the fusion device. To meet the ever increasing demands of the EAST timing system, precision time protocol (PTP) is adopted by the EAST CODAC (Control, Data Access and Communication) system to implement the prototype extension node. This paper will introduce the motivation and features of the new prototype node, the details about the slave timing module platform and test results will be described in this manuscript.

## II. CONTEXT OF THE TIMING SYSTEM

### A. Motivations for new prototype

The former timing system had a star-type topology with a central node and two local nodes. All the nodes were implemented in the PXI and FPGA industry devices, since they were located in harsh electromagnetic environment and many signal channels are required. Each timing chassis consisted of a PXI controller, a counter module and a FPGA module. The price of each commercial off-the-shelf (COTS) node was more than $ 20,000. In addition, the star-type synchronized optical topology was made up of several pairs of multi-mode fibers [1], and all the nodes distributed in experimental area were connected by these same-length fibers. The accuracy of the system synchronization depended on the difference in fiber lengths among timing nodes, and each node's delay time hinged on the route of the fiber. Therefore it's difficult for maintainers to add new timing nodes to increase the trigger signals. The motivation for this system upgrade is to set up a stable timing system that is also easy to expand, and convenient to maintain with low-cost.

### B. Synchronization protocol selection

Traditional synchronization solutions include GPS, the encoding (IRIG-B [2], Bi-Phase [3], Manchester [4] *etc.*) and the packet synchronization (NTP, SNTP). The precision time protocol (PTP) defines a protocol that allows precise clock synchronization in measurement and control systems implemented with technologies such as local computing, network communication and distributed objects [5].

The precision time protocol (PTP) is compared with other synchronization schemes as shown in Table I.

TABLE I.
COMPARISON TO OTHER PROTOCOLS

|  | GPS | NTP | PTP |
|---|---|---|---|
| Accuracy | ~ 20ns | <10ms | <1us |
| Time source | Satellite | Servers | Master clock |
| Ethernet | NO | Support | Support |
| Security | High | High | High |
| Cost | High | Low | Low |

The precision time protocol (PTP) enables heterogeneous systems that include clocks of various inherent precision, resolution, and stability to synchronize to a grandmaster clock. Moreover, it supports system-wide synchronization accuracy in the sub-microsecond range with a minimum network and local clock computing resources [5]. Custom devices which support PTP also reduce the construction costs of the EAST timing system. Compared with GPS and NTP, PTP is a good choice to meet the requirements of synchronization.

Manuscript received June 24, 2018.
This work is supported by the National Natural Science Foundation of China under Grant No.11505239.
Z. Zhang, B. Xiao, Z. Ji, Y. Wang, P. Wang are with the Department of Computer Application, Institute of Plasma Physics, Chinese Academy of Sciences, Hefei, Anhui, 230031, PR China (e-mail: zzc@ipp.ac.cn).

## III. DESIGN AND IMPLEMENTATION OF SLAVE NODE

### A. Implementation of hardware circuit

A timestamp event is generated at the time of transmission and reception of any event message. The timestamp event occurs when the message's timestamp point crosses the boundary between the node and the network [5]. As a PTP message traverses the protocol stack in a node, the timestamps are generated when the message timestamp point passes a defined point in the stack.

The point in the application layer by running PTPd on Linux, illustrated by "C" in Figure 1; in the chip kernel or media access control, illustrated by "B", or in the physical layer of the protocol stack illustrated by "A".

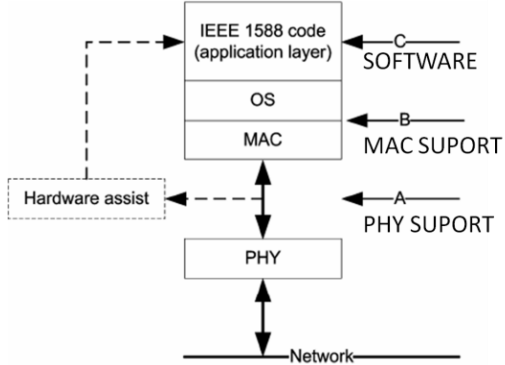

Fig. 1. Timestamp implementation model

In general, the closer this point is to the actual network connection, the smaller the timing errors introduced by fluctuations in the time taken to traverse the lower layers. The manuscript chooses "B" approach to custom the first prototype devices. The hardware design of timing system slave module uses STM32F407 and LAN8720. Cortex-M4 STM32F407 which includes a MAC802.3 controller [6] features Ethernet MAC 10/100 with IEEE 1588 v2 support. LAN8270 is a low power Ethernet PHY chip. This design uses the processor itself having timestamp function and the local clock frequency adjustment function, connects LAN8720 and STM32F407 via reduced media independent interface (RMII), converts the digital signal into two differential signal by RJ45 [7]. The hardware connection diagram is shown in Fig. 2.

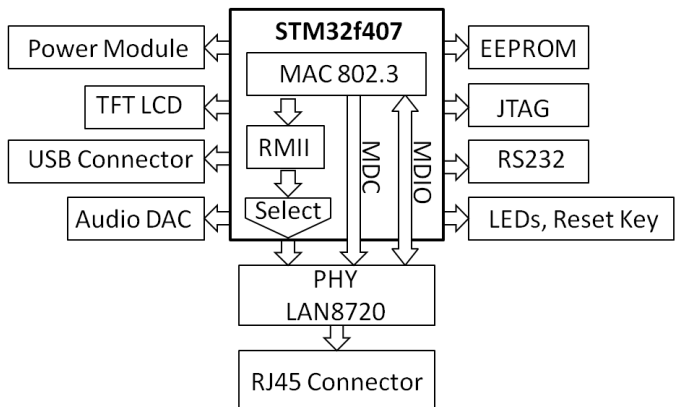

Fig. 2. Hardware block diagram of slave module

The hardware circuit includes power module, TFT LCD, RS232, *etc*. MDC (management data clock) clock line provides the timing reference for the data transfer, MDIO (management data input/output) transfer status information to/from the PHY device synchronously with the MDC clock signal.

### B. Design of Software based on PTPd v2

The LwIP (Light weight IP) is a free TCP/IP stack developed by Adam Dunkels at the Swedish institute of computer science (SICS) and licensed under the BSD license. The LwIP offers three types of API (application programming interface) [8]:
- a raw API: it is the native API used by the LwIP stack itself to interface with the different protocols.
- a Netconn API: it is a sequential API with a higher level of abstraction than the raw API.
- a socket API: it is a Berkeley-like API

The API in this prototype timing module used to build the precision time protocol with STM32F407 is the raw API. It achieves the highest performance and does not require the use of an operating system. The PTP daemon version 2 (PTPd v2) is an open source implementation of the precision time protocol (PTP) as defined by the IEEE 1588-2008 standards. The code for PTPd v2 is freely available from network.

PTPd v2 source code is grouped into a few components. The following is a block diagram of PTPd's major components, in which occlusion indicates interfaces between components.

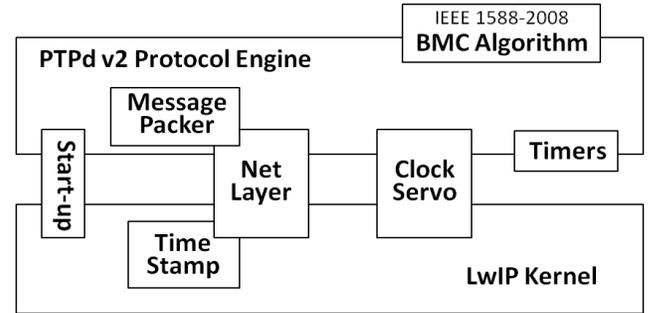

Fig. 3. Block diagram of PTPd v2's major components

The component delineations are based on the functionality, and the main functions are summarized in Table II.

TABLE II
MAIN COMPONENTS' FUNCTION

| | |
|---|---|
| PTP Engine | Implements state machine with use of P2P TC and peer delay mechanism |
| BMC Algorithm | Finds out the best master clock with IEEE 1588-2008 standard |
| Msg Packer | Gathers data into and extracts data from PTP messages |
| Net Layer | Initializes connections, sends, and receives data between PTP clocks |
| Clock Servo | Computes the offset from the M-to-S delay and S-to-M delays |
| Start-up | Sets the program's execution state, and retrieves run-time options from the user |
| Timers | Controls message exchange periodic between PTP clocks |

EAST timing system slave node main program flow chart is shown in Fig.4. The software configures SLAVE_ONLY to "TURE" during initialization, and the state machine of prototype node makes a transition from "INITIALIZING" to "SLAVE" state. The node listens to the local port to get messages. On receiving a new PTP message, the slave node records the timestamp, and extracts data information. The slave node can compute the deviation and correct error from Sync, Follw_up, and Delay_Resp messages, update the synchronization with master clock.

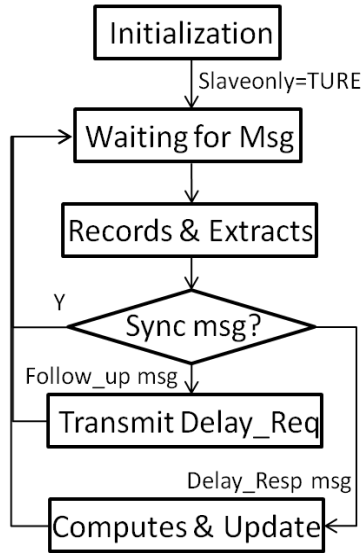

Fig. 4. EAST timing system slave node main program flow chart

### IV. Synchronization Experimental Test

The digital 1 PPS (Pulse per Second) signal is a widely used reference signal for time synchronization. The signal is a simple rectangular 1 Hz pulse, whose rising or falling edge marks the beginning of a new second [9]. In the EAST timing slave node, the pulse width is set to 10 ms. Fig. 5 shows the synchronization accuracy test platform.

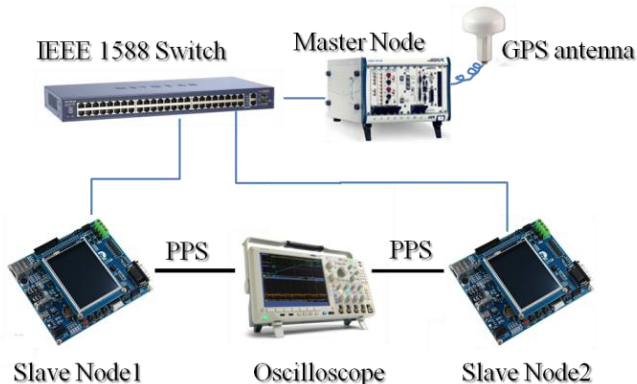

Fig. 5. Test platform between slave nodes

When a system is initialized, the node which receives the GPS signal is defined as the master node as mandated by the software process. The master node and two slave nodes (ARM STM32 platform) are connected to a switch [10]. The model of the switch is Hirschmann MAR1040 which is widely used in ITER, and the professional facility brings significantly accuracy. When the slave clock receives synchronization messages from the master clock, the slave clock gets the offset and delay value to improve synchronization accuracy. The PPS waveforms of two slave clocks are snapped by Agilent oscilloscope in Fig.6. The rising edges between the two slave PPS signals are less than 100 ns.

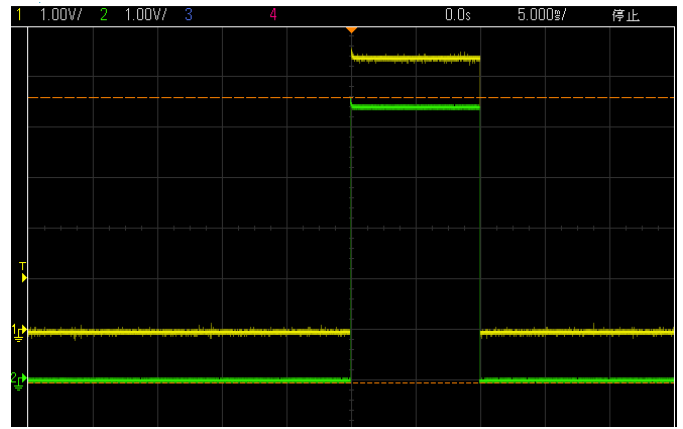

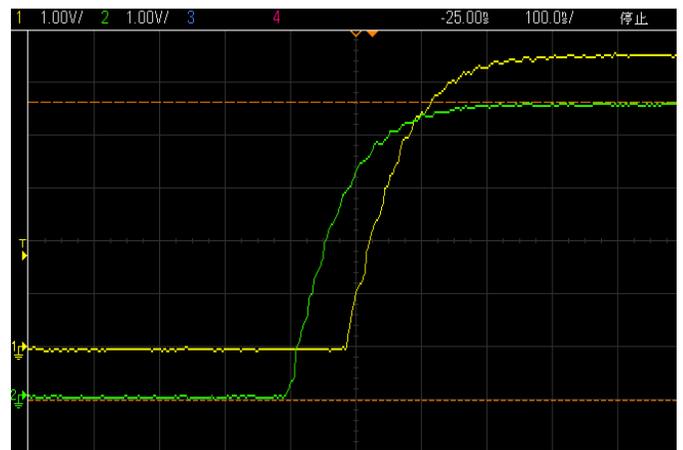

Fig. 6. Test results of slave nodes' PPS signals

### V. SUMMARY

The precision time protocol (PTP) standard is adopted by the EAST CODAC (Control, Data Access and Communication) system to implement the extended prototype timing slave node. All the nodes with PTP v2 in different places have access to the timing network by normal Ethernet cables. It's easy to expand the slave nodes. STM32F407 PPS signals are tested with synchronization accuracy less than 100 ns meet the prototype requirements. It's a good approach to reduce construction costs of the EAST timing system.

ACKNOWLEDGMENT

The author would like to thank the EAST CODAC Team for their work and help.